\documentclass[%
 reprint,
 amsmath,amssymb,
 aps,
]{revtex4-2}

\usepackage{graphicx}
\usepackage{dcolumn}
\usepackage{bm}

\usepackage{subfigure}

\usepackage{xcolor}

\begin{document}

\preprint{APS/123-QED}

\title{A novel approach for converting spatio-temporal series into complex networks} 

\author{G.Cigdem Yalcin}
\affiliation{Physics Department, Faculty of Sciences, Istanbul University, 34134, Vezneciler, Istanbul}%
\author{ M.Berk Onder}
\affiliation{Institute of Graduate Studies in Sciences, Istanbul University, 34134, Vezneciler, Istanbul}%

\date{\today}

\begin{abstract}
This study aims to offer a new perspective on complex
network representation of real-world systems. Currently, the most well-known transformation 
algorithms in the literature treat each data point in a time 
series as a node and transform the time series into a 
network. In this study, we present a new approach converting 
spatio-temporal series into a complex network. We focus on studying this transformation by grounding it in the context of physics , with the aim of adapting it to real-world problems, which often manifest as complex systems across various domains. We introduce the Gravitational Graph (GG) algorithm, which is grounded in the concept of gravitational force from fundamental physics. We consider air pollution concentrations, which represent a global environmental health risk, as an example of a complex 
environmental system, and apply the GG algorithm to 
particulate matter of 10 microns (PM10) recorded by 
21 air quality monitoring stations located in various 
regions of Istanbul, Türkiye. While the GG algorithm 
allows for the conversion of spatio-temporal series—
rather than time series—into networks, it also 
enables the analysis of the statistical properties 
and characteristics of the converted networks, 
thereby uncovering hidden relationships and 
dependencies that may not be apparent in the original 
time series.

\end{abstract}

\maketitle


\section{\label{sec:level1}Introduction }

To tackle real-world problems characterized by 
complex behaviors, it is essential to investigate 
complex interactions between the components of the 
system and its environment. Mostly real-world 
systems, whether natural or man-made, exhibit 
complex interactions and dependencies that evolve 
over time. Many physical, biological and 
technological systems, such as power grids, the 
internet, neural networks, food webs, 
transportation systems, social networks and supply 
chains inherently possess networks structure. 
In order to accurately comprehend these complex 
network systems and explore both their topological 
and functional properties, network research has 
undergone rapid and significant developments \cite{wasserman1994,newman2002,newman2010,barabasi2016,caldarelli2016}.

Transforming time series data into a complex 
network representation helps to gain a deeper  
understanding of the given system's underlying
dynamics. Therefore, well-known methods have 
been proposed for constructing a complex network 
from time series. One of the first and most 
prominent instances of transforming time series 
into complex networks involved constructing a 
network from a pseudoperiodic time series 
\cite{zhang2006}; another approach used 
\textit{visibility} between data points 
\cite{lacasa2008}; and a more recent method 
draws on electrostatics in physics 
\cite{panos2021}. Several other studies have 
contributed to this area; comprehensive reviews 
discuss a variety of these techniques—including 
but not limited to the three mentioned above—and 
their widespread applications across different 
fields \cite{zou2019, gao2021}. The main common 
features of these techniques are that each of 
them considers a single time series, treats each 
data point in the time series as one node, and 
establishes a network by determining the edges 
between these nodes according to the criteria of 
their proposed method. Although the transformation
of time series into complex networks has gained 
significant attention in the literature, in this 
letter we present a novel approach motivated by 
the existing need to transform spatio-temporal 
series containing both spatial and temporal 
attributes into a complex network.

In this letter, we present the Gravitational Graph 
(GG) algorithm that is inspired by gravitational 
force in fundamental physics. In this approach, 
the nodes of the network are defined directly as 
the real-world coordinates of geographical  
locations of the sources from which the spatio-
temporal series are obtained, while the proposed 
algorithm demonstrates how to determine the edges 
between these nodes. We demonstrate for the first 
time that gravitational force can serve as the 
basis for transforming spatio-temporal series into complex networks by identifying edges between 
nodes in a network, although gravitational force 
has previously been used in the literature to 
identify influential, attractive, or important 
nodes in complex networks 
\cite{bi2021,meng2025,chang2009}.Creating a 
complex network by transforming from a spatio-
temporal real-world data enables revealing two 
important concepts of the given system to be
discovered through complex network analysis. One 
of the key concepts is that complex network 
representation helps to capture nonlinear 
relationships between components of complex 
systems, revealing hidden relationships and 
dependencies that are not recognizable in the 
original time series. The other is to identify 
influential nodes and edges in complex networks 
that may significantly impact overall system 
behavior and correspond to critical time points or 
regions in the spatio-temporal series.

\begin{figure}[h]
\includegraphics[width=0.45\textwidth]{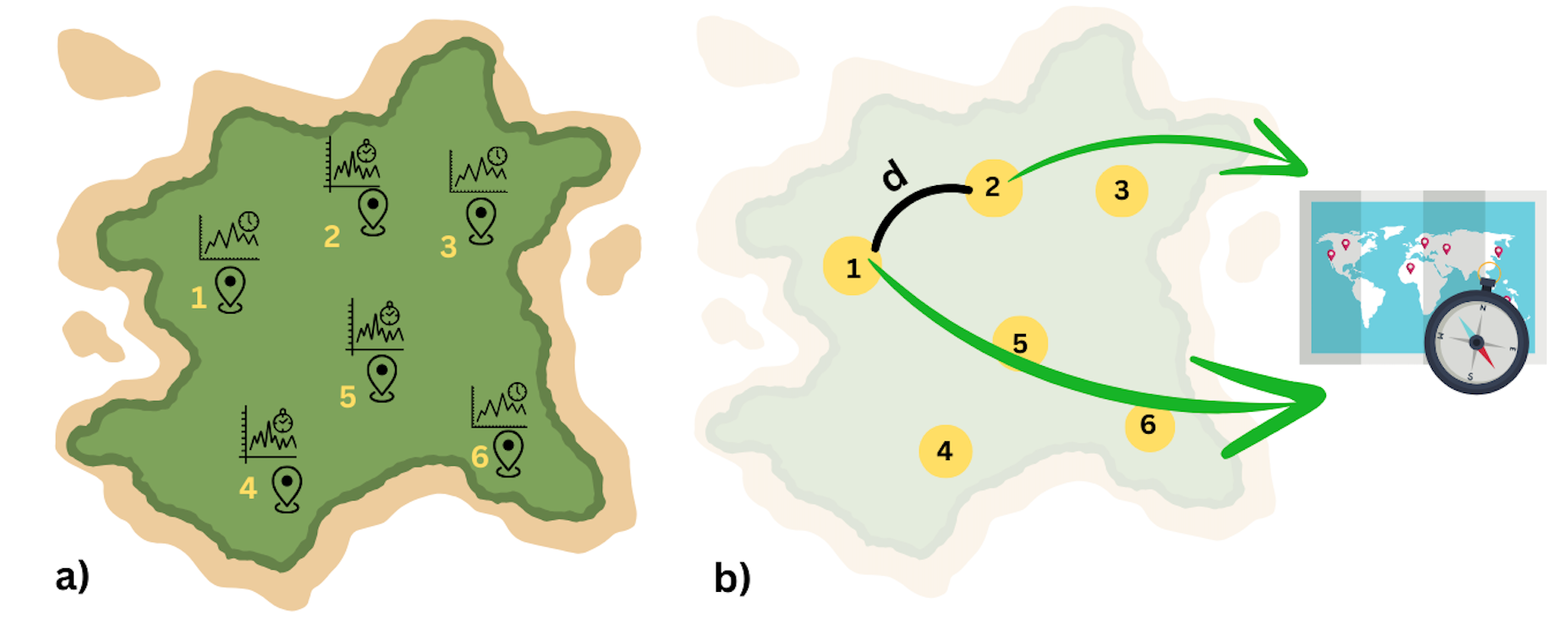}
\caption{\label{fig:representative} 
(a) The landmarks on the representative map indicate the coordinates of monitoring stations, while the graphic symbols represent the recorded spatio-temporal series at each location.
(b) The coordinates of real-world locations correspond to the nodes of a network. The distance 
$d$ represents the actual distance between the nodes.}
\end{figure}

In our approach to transforming spatio-temporal
series into a complex network, we begin by
determining the positions of the nodes of the
complex network. Fig~\ref{fig:representative}(a)
illustrates a schematic map of representative
geographical locations of data sources such as
monitoring stations or data acquisition points and
also indicates that measurement data are available
at each marked geographical location. 
Fig~\ref{fig:representative}(b) illustrates how 
each geographical location marked on the map 
corresponds to a node in the complex network 
transformed from the spatio-temporal series. Here, 
\textit{d} denotes the actual geographical 
distance between two nodes; we calculate this real-
world distance by applying the Haversine formula--
- which is widely used in many practical
applications \cite{moseli2022,maria2020}--- to the
latitude and longitude coordinates of the nodes. 
It calculates simply that
\setlength{\parskip}{0pt}
\begin{eqnarray}
\left.\
 hav(\theta)=
 sin^2(\frac{\theta}{2})%
\right.\
=\frac{1-cos(\theta)}{2}%
\label{eq:1}
\end{eqnarray}
\setlength{\parskip}{0pt}
\begin{eqnarray}
\left.\
d=
\right.\
r{archav}(hav{\theta})
\label{eq:2}
\end{eqnarray}

\begin{figure}[h]
\includegraphics[width=0.45\textwidth]{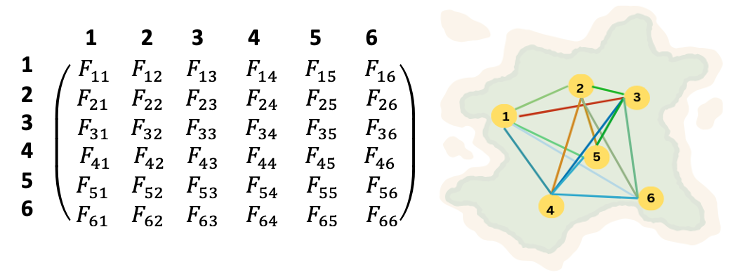}
\caption{\label{fig:matris} The elements of the adjacency matrix, calculated using the Gravitation Graph approach, correspond to the weight of the links in the weighted network.}
\end{figure}

Our proposed Gravity Graph (GG) approach generates 
an adjacency matrix, where each matrix element 
specifies the value of the edge between pairs of
nodes; for example, the $F_{12}$ matrix element 
corresponds to the edge between the \textit{node 
1} and the \textit{node 2}, resulting in a 
weighted network. This algorithm, based on the 
gravitational force formula in Eq.(\ref{eq:3}), 
considers that $m_1$ and $m_2$ correspond to the 
sum of the measured values of the data sources at 
locations labeled 1 and 2, and $d$ corresponds to 
the actual distance between these given locations. 
However, in applications of the GG approach across 
different domains, $m_1$ and $m_2$ may represent 
statistical quantities other than the sum, 
depending on the specific characteristics of the 
subject matter. As seen in the representative 
visualization in Fig.~\ref{fig:matris}, 
calculating the adjacency matrix of the spatio-
temporal series using the GG approach transforms 
it into a weighted complex network, thereby 
enabling the application of network analysis 
techniques for a more in-depth investigation. 
While a weighted network can be constructed by 
calculating edge weights from the elements of the 
adjacency matrix, an unweighted network can be 
obtained by applying a threshold to these 
elements, with the threshold determined based on 
the characteristics of the system under 
consideration.
\setlength{\parskip}{0pt}
\begin{eqnarray}
\left.\
 F=G
 \frac{m_1 m_2}{d^2}%
\right.\
\label{eq:3}
\end{eqnarray}

\begin{figure}[h]
\includegraphics[width=0.44\textwidth]{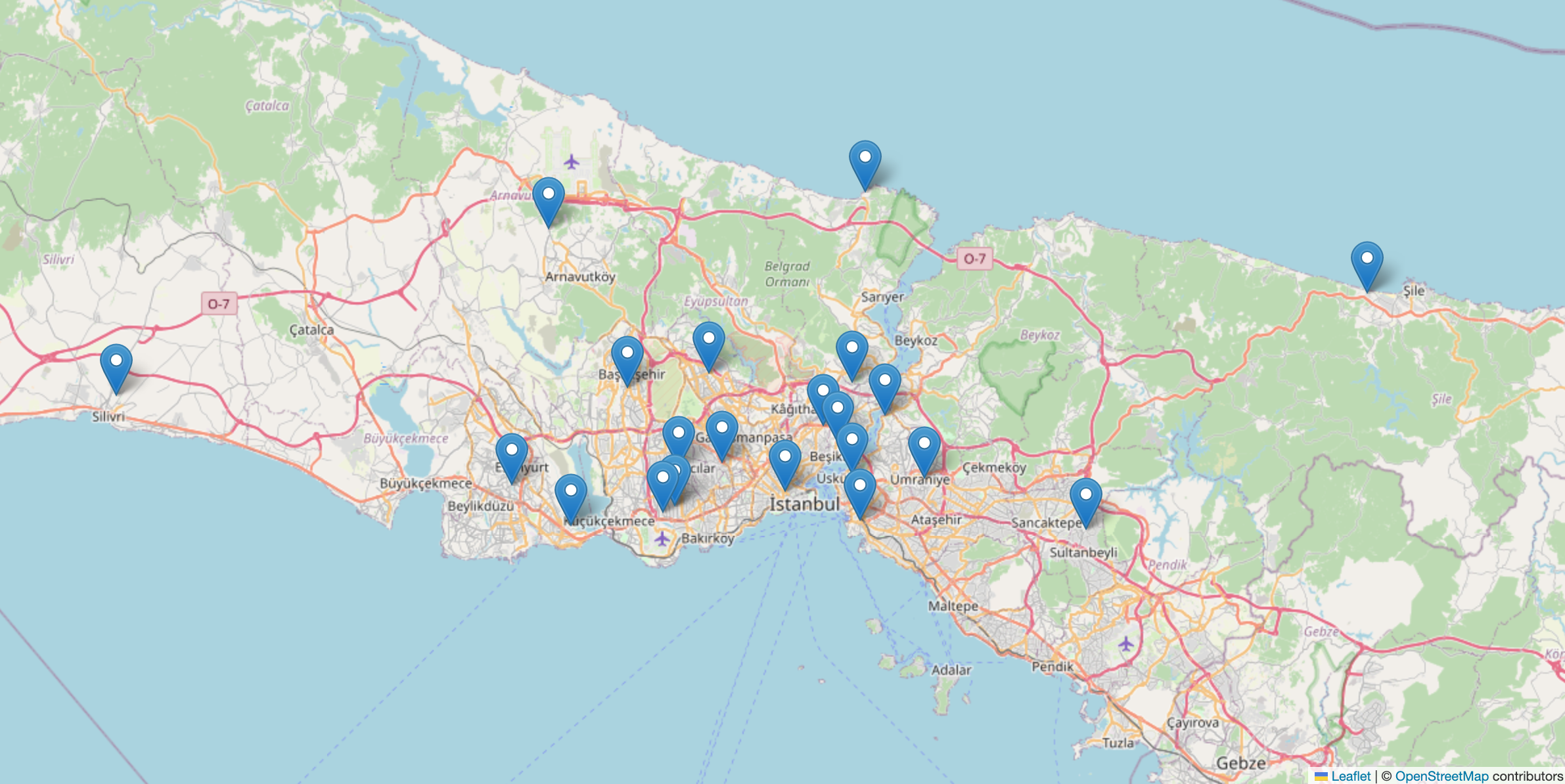}
\caption{\label{fig:map} The locations of 21 national air quality monitoring stations in various districts of Istanbul are shown on the map in real world coordinates (latitude-longitude). }
\end{figure}

To exemplify the implementation of the GG 
algorithm on real-world spatio-temporal series, we 
consider air pollution as a complex environmental 
system. Air pollution has been extensively 
examined as a complex system, employing 
methodologies that go beyond traditional 
analytical techniques \cite{williams2020}. We 
selected this topic as an example due to its 
prominence in complex systems research and 
because, despite technological and environmental 
advances, air pollution continues to be the 
leading environmental health risk in Europe and 
a major global issue 
\cite{european2024,european2025}. We focus on 
PM10 concentration, which refers to particulate 
matter with a diameter of 10 micrometers or 
less; as it is respirable and can cause 
serious health problems \cite{gial2025}. The 
publicly available daily average PM10 
concentration data \cite{pm10data} used in this 
study were measured at national air quality 
monitoring stations in 21 locations across 
Istanbul, Türkiye, covering a period of six 
years from 2019 to 2024, as shown on the map in 
Fig.~\ref{fig:map}. With its unique location 
spanning two continents and a population 
exceeding 16 million, Istanbul experiences a 
high and variable pollution load driven by 
emissions from transportation, industry, 
residential heating, and ongoing urban expansion.

\begin{figure}[h]
\includegraphics[width=0.42\textwidth]{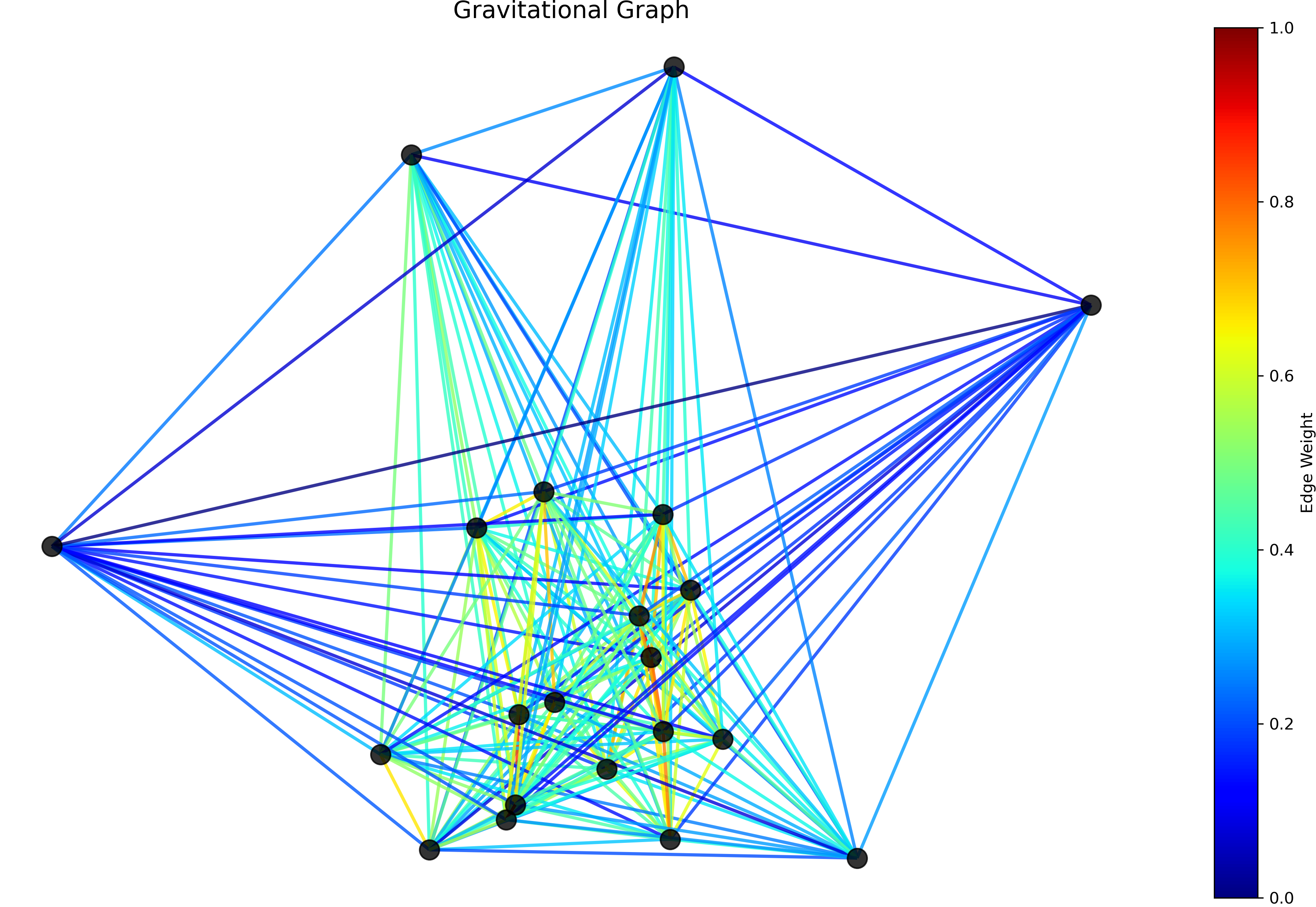}
\caption{\label{fig:gravitational_graph} Weighted complex network generated by the GG method.}
\end{figure}

In this application, we transformed the spatio-
temporal air quality data of Istanbul into a 
complex network using the GG algorithm. In
Eq.~\ref{eq:3}, the mass quantities $m_1$ and 
$m_2$ correspond to the sums of the daily 
average measured values of air pollution (PM10) 
at the two respective stations. Summing the data 
points of an air pollution time series provides 
an indication of the total exposure to pollution 
over the given time period. This cumulative  
exposure, derived from aggregating daily 
averages, helps assess the overall pollution 
burden experienced by an area, which is 
essential for evaluating long-term health risks 
and environmental impacts. For distance $d$, we 
calculate the actual distance between the two 
stations based on their latitude and longitude 
values using the Haversine formula as in Eq.(\ref{eq:1}). Thus, by using the GG algorithm, 
we calculate the elements of the adjacency 
matrix corresponding to the edge between each 
pair of nodes in a network and construct the
matrix. As a result, we create the weighted complex network shown in Fig.~\ref{fig:gravitational_graph}.

\begin{figure*}
\setkeys{Gin}{width=1\linewidth}
\begin{minipage}[t]{0.24\textwidth}
\includegraphics{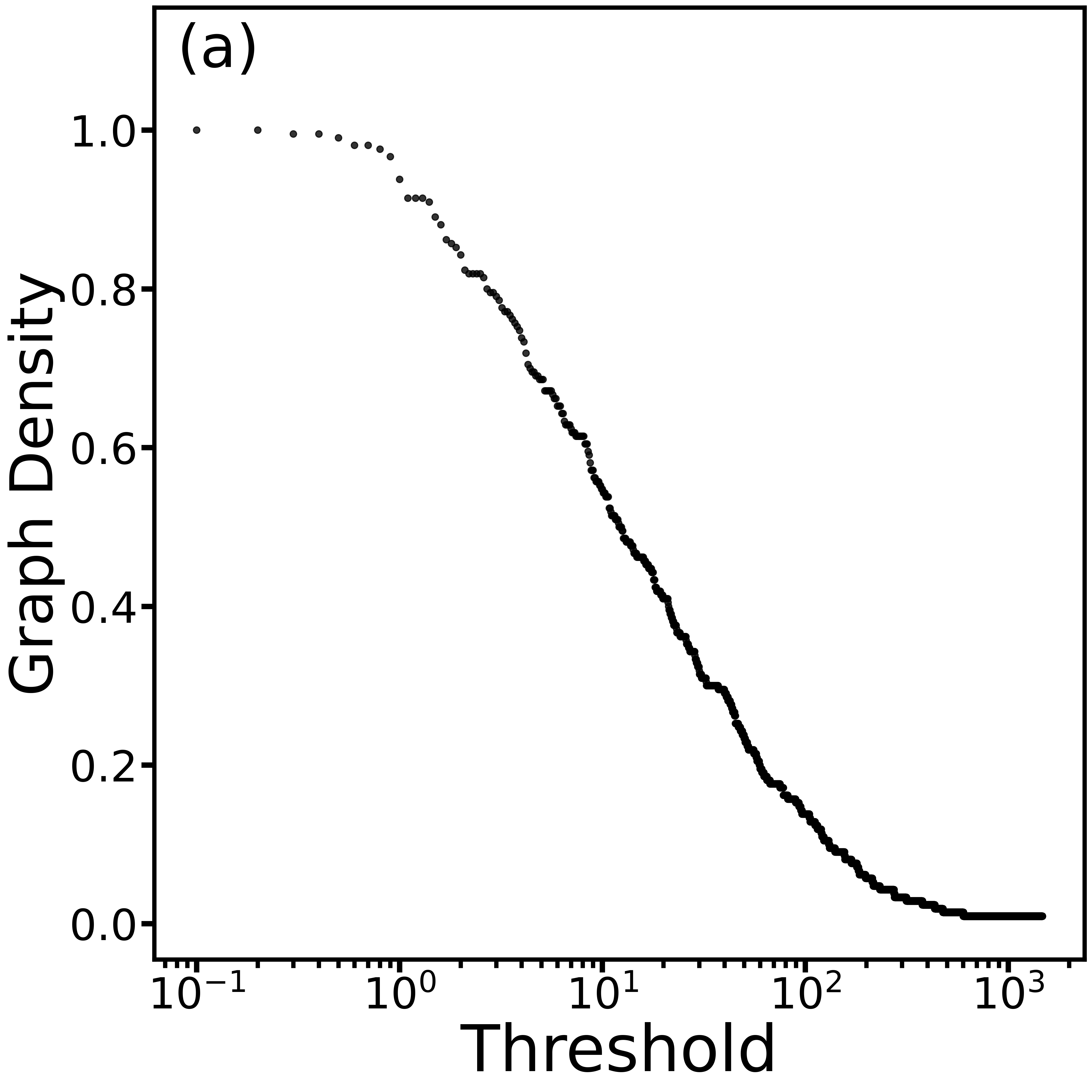}
\label{fig:a}
\end{minipage}
\hfill
\begin{minipage}[t]{0.24\textwidth}
\includegraphics{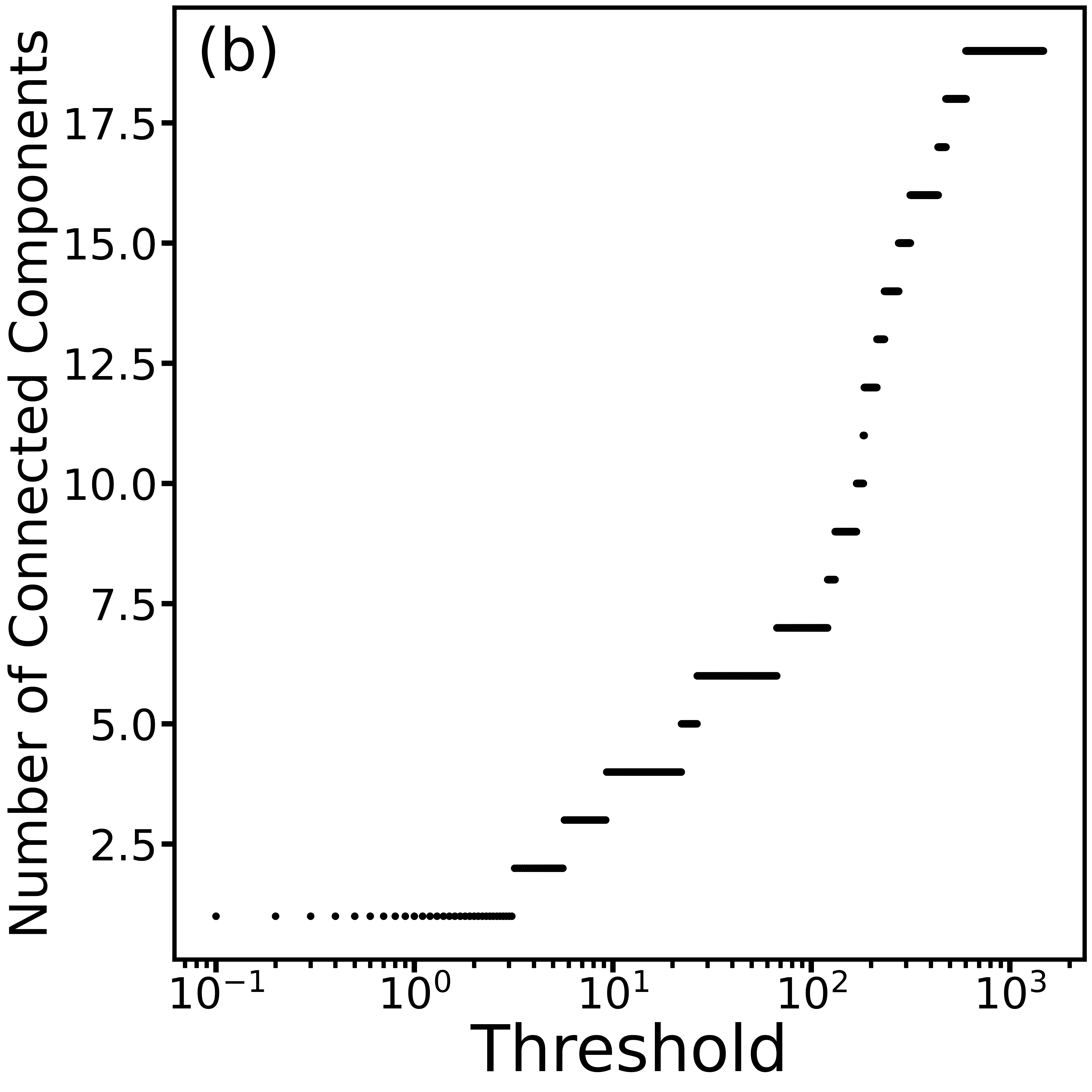}
\label{fig:b}
\end{minipage}
\hfill
\begin{minipage}[t]{0.24\textwidth}
\includegraphics{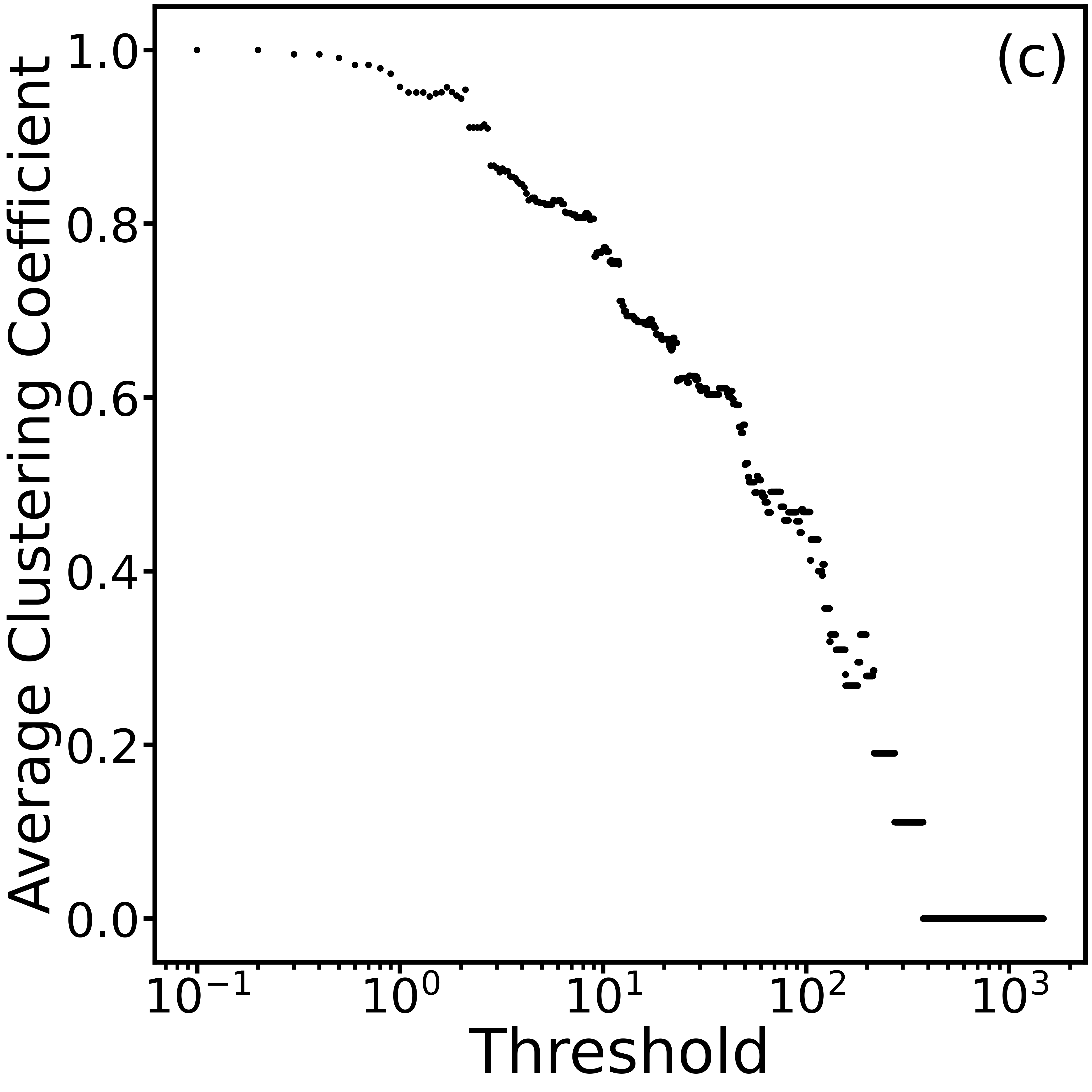}
\label{fig:c}
\end{minipage}
\hfill
\begin{minipage}[t]{0.24\textwidth}
\includegraphics{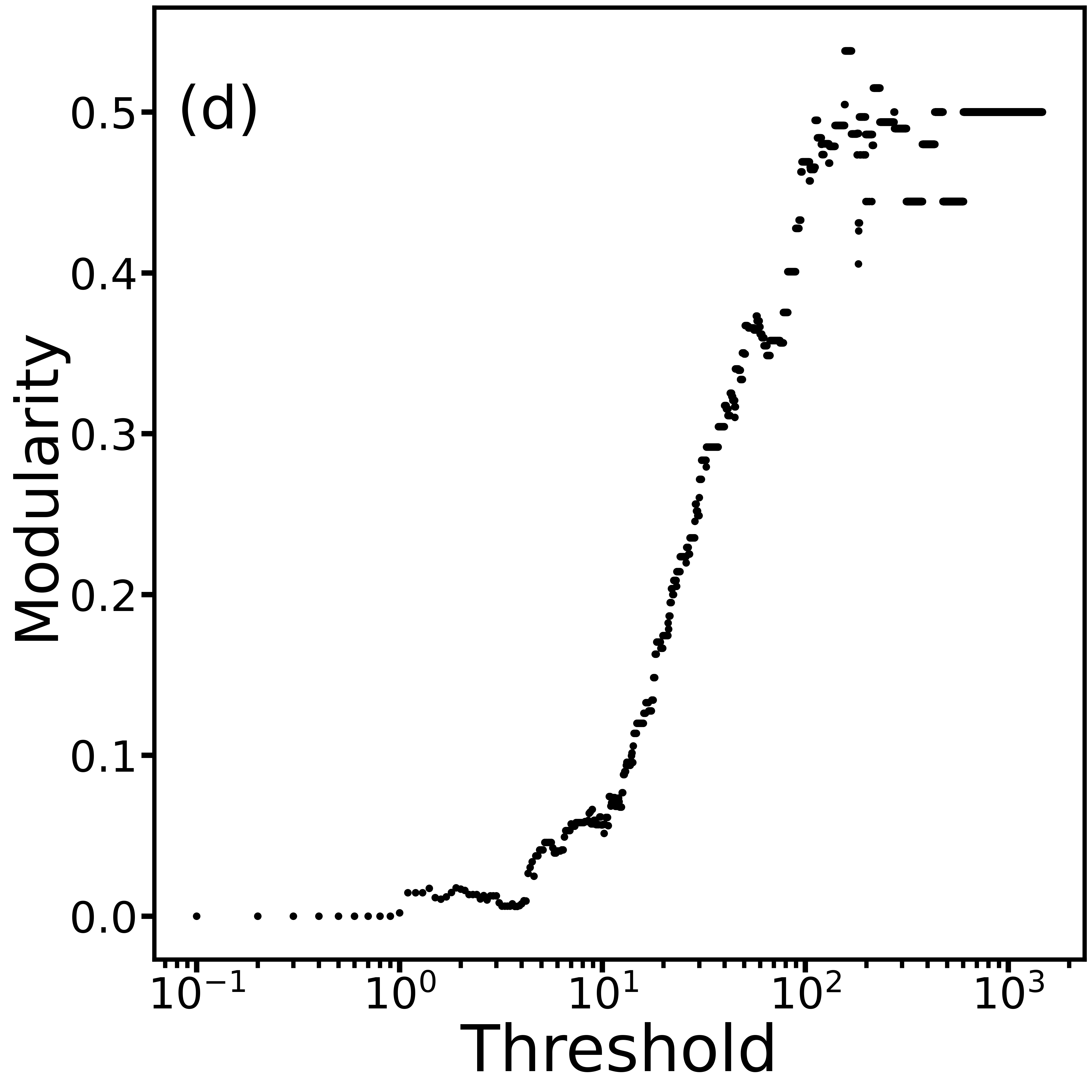} 
\label{fig:d}
\end{minipage}
\\
\begin{minipage}[t]{0.24\textwidth}
\includegraphics{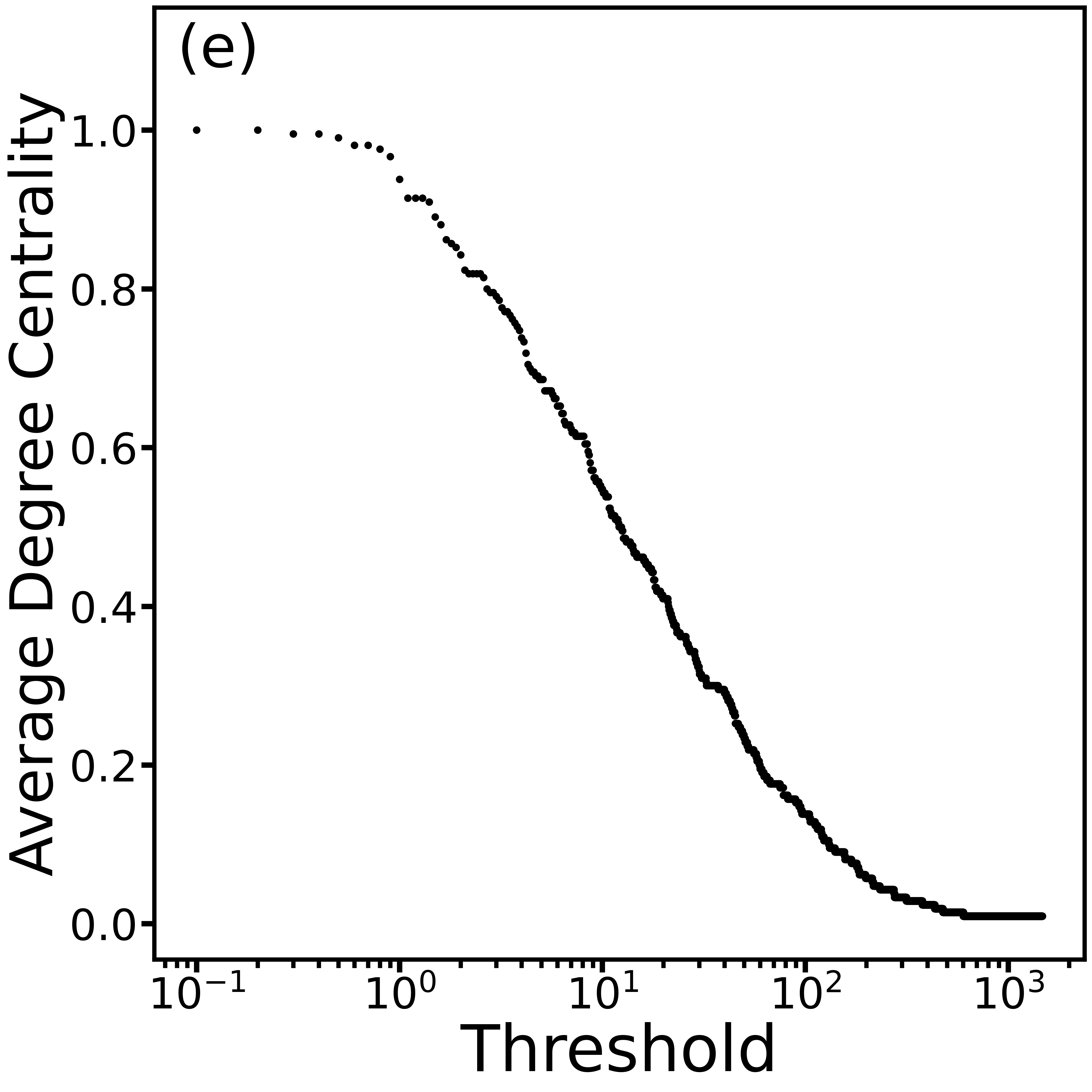}
\label{fig:e}
\end{minipage}
\hfill
\begin{minipage}[t]{0.24\textwidth}
\includegraphics{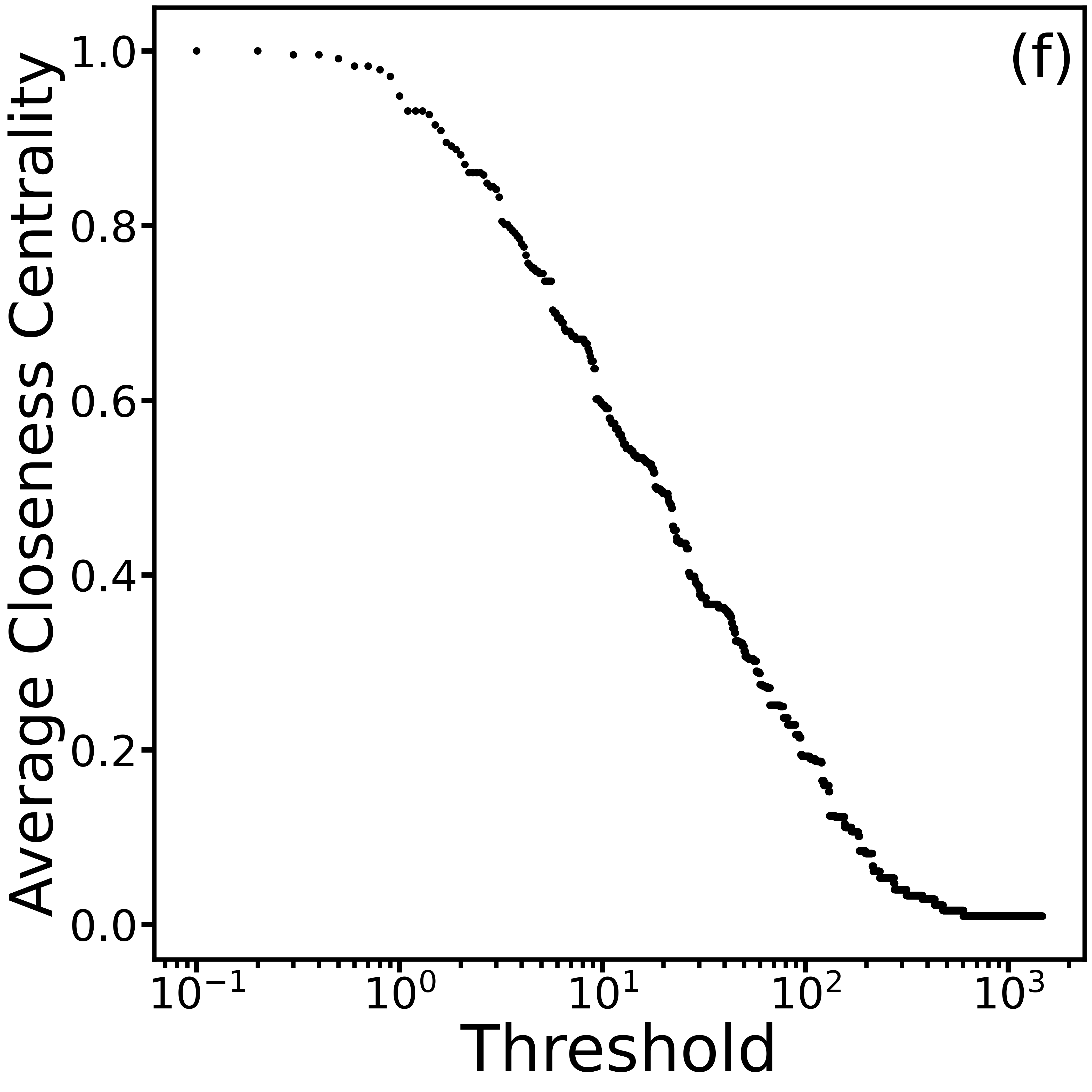}
\label{fig:f}
\end{minipage}
\hfill
\begin{minipage}[t]{0.24\textwidth}
\includegraphics{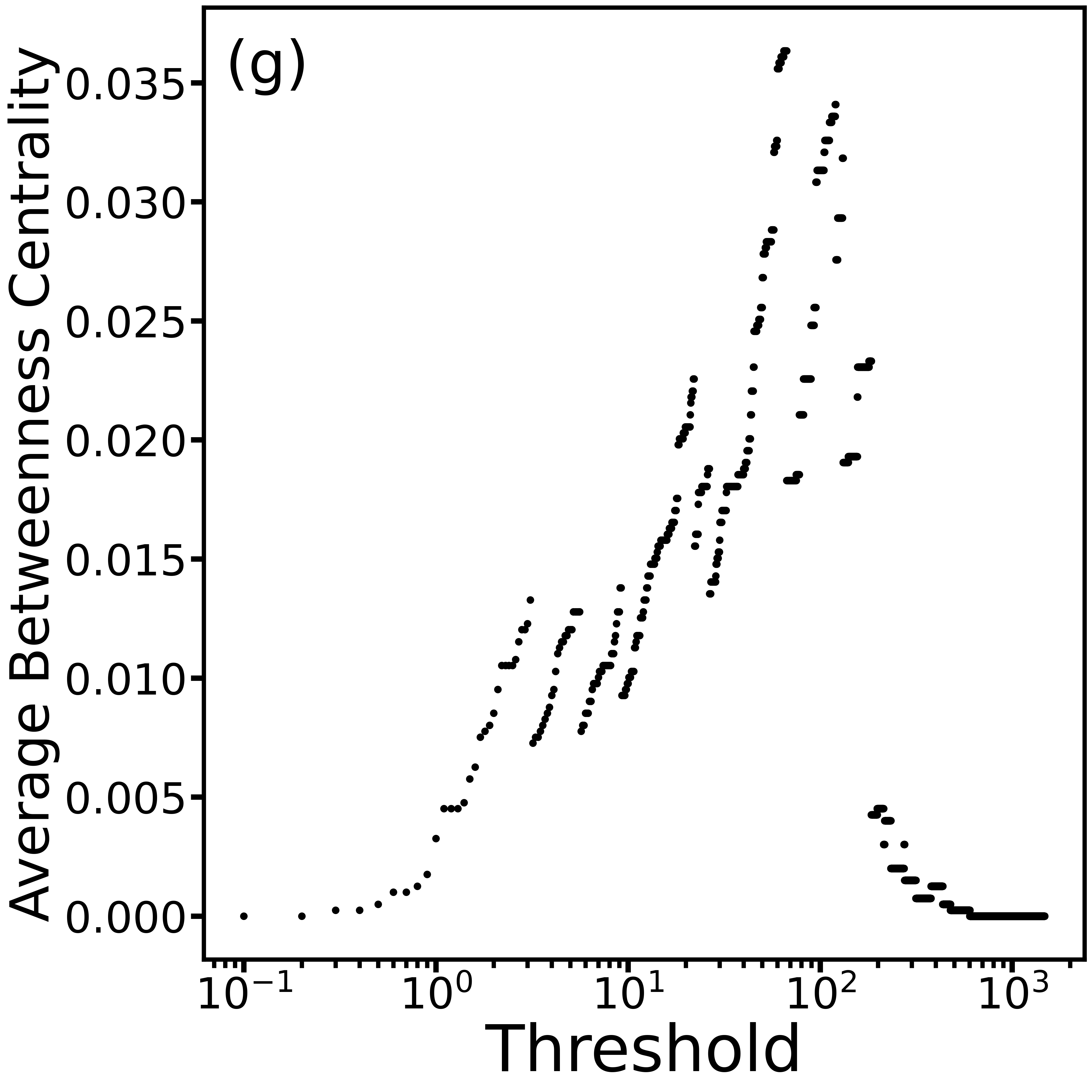}
\label{fig:g}
\end{minipage}
\hfill
\begin{minipage}[t]{0.24\textwidth}
\includegraphics{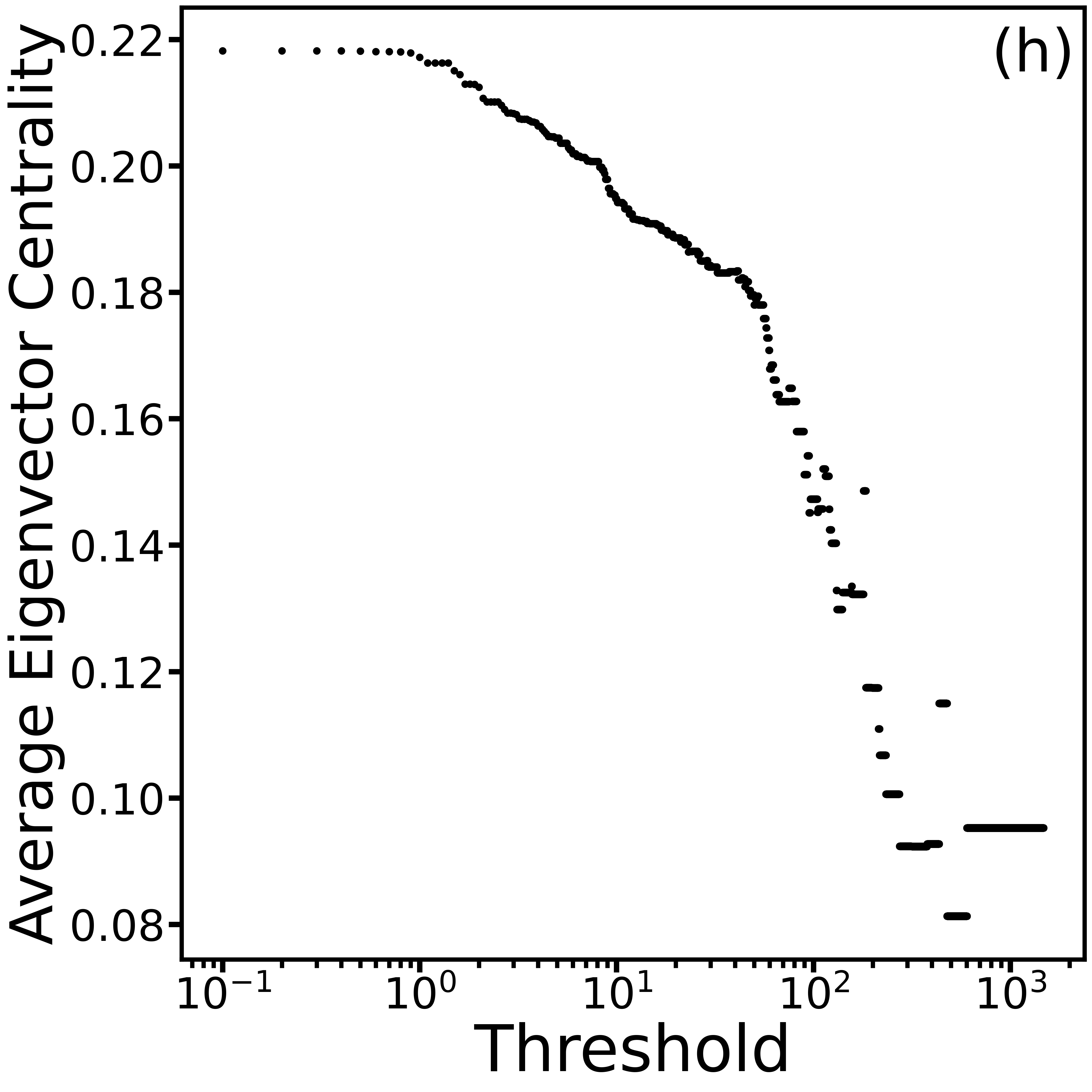} 
\label{fig:h}
\end{minipage}
\caption{Variation of key network properties in complex networks generated by applying different thresholds to the adjacency matrix.
a)Graph density,
b)Number of connected components,
c)Average clustering coefficient,
d)Modularity, e)Average degree
centrality, f)Average closeness
centrality, g)Average betweenness
centrality, h)Average eigenvector
centrality.}
\label{fig:wide8}
\end{figure*}

When setting a threshold to be applied to the 
adjacency matrix for creating an unweighted 
network, it is appropriate to determine the 
threshold according to the specific 
characteristics of the subject under study. 
Here, rather than applying a single fixed 
threshold to the elements of the adjacency 
matrix to construct an unweighted network, we 
systematically applied thresholds ranging from 0 
to 1 to the adjacency matrix. By applying 
different thresholds to the adjacency matrix, we 
obtain a series of distinct complex networks. 
The various properties of these networks were 
computed and systematically compared to one 
another. For the networks created as the 
threshold is varied, we calculate key network 
properties \cite{freeman1978,bonacich1987,scott2000,bilke2001,eriksen2003,nishikawa2003,newman2004,newman2005,boccaletti2006,brandes2008,sayama2015,barabasi2016} at three scales: microscopic, node-level 
measures (clustering coefficient, centrality 
measures); macroscopic, whole-network-level 
measures (graph density); and mesoscopic 
measures (modularity), which emerge at an 
intermediate scale between microscopic and 
macroscopic level. To study how network
structure changes as the threshold increases,
key properties were analyzed, as illustrated in
Fig.~\ref{fig:wide8}(a–h).


The graph density $D$ is the ratio of the actual number of edges in the graph to the maximum possible number of edges and it is
calculated as $D$ = $\frac{2E}{N(N-1)}$ and indicates how densely or sparsely connected the networks are.
As shown in Fig.~\ref{fig:wide8}(a), increasing the thresholds applied to the adjacency matrix leads to a decrease in the 
graph density of the resulting networks.

The number of connected components in a 
network refers to the number of subgrapghs. 
Depth-First Search (DFS) or Breadth-First
Search (BFS) are among the most widely used graph 
algorithms to efficiently count the number of
connected components. In Fig.~\ref{fig:wide8}(b) 
shows that the change in the number of components
in the networks formed with distinct thresholds and it follows a pattern 
that reflects an increase in the number of fragmentations in networks formed at higher thresholds.

The clustering coefficient $C_i$ of node
$i$,
$C_i = \frac{2E_i} {k_i (k_i - 1)}$ is
defined
as the number of triangles $E_i$ of
involving
node $i$ is divided by the number of the
total
triangles $k_i(k_i-1)$ where $k_i$ is 
the degree of node $i$. The average
clustering coefficient
of the entire network
$ C = \frac{1}{N}\sum_{i=1}^{N} C_i $
measures how connected a node's neighbors are to 
each other, to give an idea of the local 
interconnectedness of the network. 
According to the results in Fig.~\ref{fig:wide8}
(c), as the thresholds increases, the number 
of triangle formations in the resulting networks 
decreases, and hence the local connectivity.

Modularity is an indicator of how a
network structure is organized into 
multiple communities such as groups,
clusters. It is defined as \(Q=\frac{1}
{4L} \sum_{ij}
\left(A_{ij} - \frac{k_i k_j}{2L} 
\right)
\delta(c_i,c_j) \) where \( A_{ij} \) 
is the adjacency matrix, \( k_i \)
and \( k_j \) are degrees of node $i$ and 
$j$, \( L\) is the total number of
edges, and \( \delta(c_i, c_j) \) is 1
if nodes \( i \) and \( j \) are in
the same community, otherwise 0.
The results in Fig.~\ref{fig:wide8}(d)
reveal that modularity rises sharply 
with increasing threshold, indicating 
that the emergence of well-defined 
subgroups is more likely to occur in networks generated with higher thresholds.

Degree centrality is defined as $C_D(i) = 
\frac{d(i)}
{N-1}$ where $d(i)$ is the degree of node
$i$ and  $N$ is the total number of nodes in
the network. The average degree centrality is derived from the individual degree centralities:
$\bar{C}_D=\frac{1}
{N}\sum_{i=1}^{N}C_D(i)$ which is
essentially the normalized mean degree
across all nodes. Fig.~\ref{fig:wide8}(e) presents 
that in networks formed at low thresholds, higher 
average degree centrality indicates that the 
network is more connected and dense. However, in 
networks at increasing thresholds, as weak links 
are eliminated, the network becomes more 
fragmented and average degree centrality decreases.


The closeness centrality measures how
efficiently information can be spread throughout the 
network based on the average shortest
path length between nodes. It is calculated as $C_C(i) = 
\frac{(N-1)}{\sum_{j} d(i, j)}$, where $C_C(i)$ is the
closeness centrality of node $i$, $N$ is the total number
of nodes. The average closeness centrality of a
network is given as $\bar{C}_C = \frac{1}{N}\sum_{i=1}^{N} C_C(i)$.
Fig.~\ref{fig:wide8}(f) indicates that in the networks
resulting from applying higher thresholds, the average
closeness centrality decreases.

The betweenness centrality of a node $i$ is defined as: 
$C_B(i) = \sum_{\substack{j \neq
i ,k \neq i, j \neq k}} \frac{\sigma_{jk}(i)}{\sigma_{jk}}$
where $\sigma_{jk}$ is
the total number of shortest paths between nodes $j$ and $k$.
$\sigma_{jk}(i)$ is the number of those shortest paths that
pass through node $i$. Betweenness centrality at node-level
acts as a bridge along the shortest path between two other
node in the network's flow, while average betweenness
centrality of the network is given by $\bar{C}_B =
\frac{1}{N} \sum_{i=1}^{N} C_B(i)$, provides an overall
measure of how much nodes mediate, on average, along the
network's shortest paths. Fig.~\ref{fig:wide8}(g) shows
that the average betweenness centrality takes higher values in networks created at medium thresholds, while it takes lower values in networks formed at both low and high thresholds.
\setlength{\parskip}{0pt}

Eigenvector centrality quantifies a node’s influence by considering not only the number of its connections but also the significance of the nodes to which it is connected. The eigenvector $C_E$ corresponds to the $\lambda$ largest eigenvalue of the adjacency matrix $A$.
$\mathbf{A} \mathbf{C}_E = \lambda \mathbf{C}_E$,
This eigenvector gives the centrality scores of all nodes at once. However this component-wise formula $C_E(i) = \frac{1}{\lambda}
\sum_{j \in N(i)} A_{ij} C_E(j)$ just the expanded form of the matrix equation which is useful for describing the importance of flows between important neighbors of the recursive structure, not
for repeated calculation. The average eigenvector centrality $\bar{C}_E
= \frac{1}{N} \sum_{j=1}^{N} C_E(i)$,  gives a single number summarizing how globally central the nodes are, on average. Fig.~\ref{fig:wide8}(h) shows the change in eigenvector centrality measurements in networks created with increasing thresholds, indicating that it gradually decreases until a certain threshold and then starts to decrease faster and in wider ranges.

We also perform a dimensional analysis to demonstrate the consistency of the spatio-temporal air pollution data with the base equation of the GG approach. Gravitational force, given in Eq.~(\ref{eq:3}) where $G$ is the gravitational constant with units of $m^3 
kg^{-1}s^{-2}$, $m_1$ and $m_2$ have units of kg, 
$d$ is the distance in meters and $F$ has the 
expected units of Newtons $(kg.m/s^2)$. Noting that 
mass is defined as $m = \rho V$, where $\rho$ 
is density $(kg/m^3)$ and $V$ is the volume $(m^3)$, 
the rewritten force equation is $F = G 
\frac{(\rho_1 
V_1)(\rho_2 V_2)}{d^2}$. In our application, the air 
pollution data is provided in terms of density with 
units of  $\mu g/m^3$. Since this represents a scaled 
version of mass density, the physical units remain 
unchanged. Therefore, replacing mass with the 
product of density and volume in the model preserves
dimensional consistency and validates the physical 
foundation of the GG approach for analyzing spatio-temporal air pollution data.  
\setlength{\parskip}{0pt}

In summary, we introduced the Gravitational Graph (GG) approach as a novel method to convert spatio-temporal series into complex networks. Unlike existing time series transformation techniques, the GG approach specifically targets spatio-temporal series, thereby extending the applicability of network-based analyses to more realistic and complex real-world systems. Once the network is constructed by obtaining the adjacency matrix from spatio-temporal series through the GG framework, standard network analysis tools can be employed. Thus network analysis can uncover structural and dynamic features often hidden in the raw time series, enabling deeper insights into the underlying system behavior.

We implemented the GG approach on air quality spatio-temporal series, specifically PM10, recorded in Istanbul, demonstrating that this transformation enables network-based analysis of such spatio-temporal series. The spatial information of the air quality monitoring stations—specifically latitude and longitude coordinates—defined the nodes, while the PM10 data recorded at these locations are used to construct edges between nodes using the GG algorithm. This process yielded an adjacency matrix and, consequently, a weighted network. To explore different network structures, we constructed unweighted networks by applying threshold values between 0 and 1 to the adjacency matrix elements. For each of resulting networks, we computed various network properties that provide information at the microscopic, macroscopic, and mesoscopic levels. Thus, we compared the unweighted networks obtained under different thresholds and analyzed how their properties change depending on the threshold applied. On the other hand, the GG algorithm may also detect geographical influences by identifying nodes with strong or weak connections within the adjacency matrix, which represents the calculated relationships between these nodes. For example, higher air pollution measurements were observed in monitoring stations located in complex urban areas, whereas significantly lower values appeared at coastal stations, highlighting the geographical effect on air quality. Consequently, this application paves the way for more detailed investigations, such as adding extra parameters, analyzing additional network properties and selecting specific thresholds for air pollution, which will be valuable for future air quality studies. This work contributes to the understanding of real-world complex systems with spatio-temporal series through their representation as complex networks.

\textit{Acknowledgments-} G.C.Y. thanks to Istanbul University Scientific Research
Projects No: 32990. The data used in this study were obtained from the relevant departments of the Ministry of Environment, Urbanization and Climate Change of the Republic of Türkiye. The map Fig.3 was created using OpenStreetMap www.openstreetmap.org 


\nocite{*}

\bibliography{apssamp}


\end{document}